\documentclass[namedreferences]{solarphysics}

\usepackage[hyperref,optionalrh,solaromanenum]{spr-sola-addons} 
\usepackage{graphicx}                    
\usepackage{amssymb}                    
\usepackage{color}                       
\usepackage{amsbsy}

\usepackage[english]{babel}

\newcommand{\adv}      {{\it Adv. Space Res.}}

\newcommand{\aap}      {{\it Astron. Astrophys.}}

\newcommand{\apj}      {{\it Astrophys. J.}}

\newcommand{\jgr}      {{\it J. Geophys. Res.}}

\newcommand{\solphys}  {{\it Solar Phys.}}

\begin{document}

\begin{article}

\begin{opening}
\title{Validity of NLFFF Optimization Reconstruction}


\author[addressref=aff1,corref,email={rud@iszf.irk.ru}]{\inits{G.V.}\fnm{G. V.}~\lnm{Rudenko}\orcid{0000-0003-0789-1574}}

\author[addressref=aff1,corref,email={dmitrien@iszf.irk.ru}]{\inits{I.S.}\fnm{I.S.}~\lnm{Dmitrienko}}
\address[id=aff1]{Institute of Solar-Terrestrial Physics SB RAS, Lermontov St. 126,
Irkutsk 664033, Russia}

\runningauthor{G.V. Rudenko and I.S. Dmitrienko} 
\runningtitle{Validity of NLFFF Optimization Reconstruction}

\begin{abstract}
We evaluate validity of NLFFF extrapolation performed with Optimization class (OPTI) codes. While explaining
inevitable for OPTI partial non-solenoidality  caused by the gas pressure notable role in pressure balance at
photospheric heights and by mathematical aspects related to optimization and BVP (boundary value problem), we
justify elimination of the non-solenoidal component (postprocessing) from the OPTI result obtained. In essence,
postprocessing converts the entire non-solenoidal part into a solenoidal force part, which possibly reflects
factual deviation of magnetic field from its force-free pproximation on the photosphere and in the solar corona.
Two forms of postprocessing have been analyzed in this paper. Postprocessing I eliminates the non-solenoidal
component without changing transverse field at the measurement level, and Postprocessing II leaves the field
normal component unchanged.  Extrapolation, postprocessing, and then comparison of metric and energy 
characteristics are performed over AR 11158 active region for a small fragment of its evolution containing the
February X-class flare. Our version of OPTI code showed that free energy decreased by $\sim  10^{32}$ erg within 1
hour, which corresponds to theoretical estimations of the flare-caused magnetic energy loss. This result differs
significantly from the one in  \cite{Sun}. 
Therefore, we also comment on some features of our OPTI code implementation, which may cause significant
differences between our results and those obtained using the \cite{Wieg4} version of OPTI code in study by
\cite{Sun}.

\end{abstract}
\keywords{The Sun, Magnetic field, Vector magnetograms, NLFF extrapolation, Free energy,  solenoidal fields}
\end{opening}
\section{Introduction} 
As it is commonly believed, extreme solar activity is provided by magnetic energy of active regions
\citep{Forbes}. According to Thomson's (or Dirichlet's) theorem, e.g., \cite{Lawrence}, any solenoidal field can
be decomposed into the current and potential parts. The energy of the latter corresponds to the minimum possible
under equal conditions for the normal component at the region boundary. For this reason, only energy of the
current component (free energy) is responsible for the magnetic region ability to release energy, and can be used
for predicting solar flares \cite{Barnes_2016}.

Obtaining magnetic structures and their energy characteristics from measurement data accessible at photospheric
heights is only possible if approximation of the force-free field is used. This approximation is applicable for
the upper corona due to low $\beta$ – the ratio of gas-kinetic pressure (exponentially falling with height) to
magnetic pressure (falling slower).  At photospheric heights, parameter $\beta$ can be conventionally considered
low only in regions with rather high field module $|{\bf B}|$. In regions with lower $|{\bf B}|$, parameter
$\beta$ can take values of order of $1$ and greater. Because it is a strong field that generates the main
structure of the coronal magnetic field, we can notionally use the force-free approximation. The key point for our
work will exactly be the fact that getting a field model of a real active region from the force-free
approximation, we will inevitably face disagreement with the real magnetic field containing a certain force
component. In particular, free energy of the force-free approximation will also differ from free energy of a real
solenoidal field with the force component.

In practical extrapolation, many methods are developed from the force-free approximation and applied (see, e.g. 
\citep{Schrijver_2006,Metcalf}). In this paper we examine the code based on the optimization method
(OPTI), \cite{Wheatland}, one of the most used in magnetography applications of active regions. The expertise is of
fundamental importance in the context of significant non-solenoidality of numerical OPTI results discussed in
literature  \citep{Valory_2013,Mastrano_2018}. Notable non-solenoidality of OPTI results is quite natural and  is
associated with two factors. First, the OPTI algorithm implements the search for solution in the functional space
of arbitrary non-solenoidal fields, on equal terms minimizing both Lorenz's power and divergence.
If we solve a problem with boundary values (BVs) providing a force-free result, the OPTI algorithm (as known from
the \cite{Low} analytic force-free models) provides a smallness of the non-solenoidal component at the level of
numerical errors. In case there is no force-free solution satisfying the BVP, it is natural to expect that the
algorithm will lead to occurrence of the finite Lorenz force and divergence in the solution. Secondly, the BVP
specificity for OPTI (a full set of the field boundary components $B_x,B_y,B_z$ is specified) concedes the absence
of any force-free solution. In fact, we can cite the following example. If it is a force field, then quite natural
are cases when the normal field on the solar surface is zero with the nonzero component of the current normal
component. In this case, if we try to define boundary conditions for a force-free field using these data, we come
to a non-realistic infinite value of the force-free parameter $\alpha$. That is, for this case, it is notionally
impossible to find a suitable force-free field, and we have the right to extend this statement to any real
photospheric field. Thus, in OPTI solutions, we will always have the force and non-solenoidal components
consistent with the results by \citep{Valory_2013,Mastrano_2018}. Desirable  estimates of non-solenoidality for
other extrapolation techniques considered in \cite{Mastrano_2018}, apparently, can also be explained in terms of
their features and specifics of their BVP. For example, methods that belong to Grad—Rubin CFIT, XTRAPOL and FEMQ
class from \cite{Mastrano_2018} are   realized (exactly or almost exactly)   in the class of solenoidal fields and
this is first. Second, their BVP is defined by the pair $B_n$ and $\alpha(B_n>0)$. These two moments naturally
lead to force-free solutions, as shown in \cite{Mastrano_2018}. So, we have two basic differences in two
approaches to magnetic field extrapolation. When trying to save accurate  BVs of all field components, the OPTI
technique  leads to an approximate result with the force and non-solenoidal components. In the second approach, we
get some force-free and solenoidal result, while giving away accurate reproduction of the boundary field that we
obtained in measurements. Since the field on the photosphere is different from the force-free field, the
force-free parameter $\alpha $, strictly speaking, does not make sense in its exact meaning, i.e. we to some
extent select boundary values of the field.  In this situation, no one can say results of which of the two
approaches are more realistic. It is likely that in an effort to obtain an accurate force-free field, we can move
further away from the real field, which has a noticeable force component. In order to clarify the OPTI problem
highlighted in studies by \citep{Valory_2013,Mastrano_2018}, we implement one of the two types of postprocessing.
Both types of postprocessing completely remove non-solenoidality from our solution, and convert it into an
additional force component. Postprocessing I leaves the transverse field unchanged at the bottom boundary of the
extrapolation box using the divergence cleaner algorithm by \cite{Valory_2013}, Formula (B.4). On the contrary,
Postprocessing II  leaves the normal field unchanged by adding a compensating potential component to the field of
Postprocessing I. After any postprocessing, final solution includes a certain non-zero force component. This force
component, if not taken as a mistake, is likely to indicate the above-mentioned deviation of the real solar
magnetic field from the force-free field.

Our research is conducted on a well-studied part of AR 11158 evolution that contains the X-class flare on February
15, 2011 01:44 UT. Extrapolation is carried out with our own version of OPTI technical implementation. We compare
the AR 11158 energy characteristics within the selected time interval and the overall structure of field  lines
with the results of \cite{Sun}. It is important that our results before and after any postprocessing show
approximately the same $\sim 10^{32}$ erg drop in free energy during one hour after the flare onset. This matches
the estimates of the required reserve energy for X-class events \citep{Hudson,Bleybel}. Such a great discrepancy
with the result by \cite{Sun} seems to be caused by the fundamental difference in some aspects of our OPTI
implementation that we point out here. 

For results before and after one of the two postprocessings, we perform a detailed analysis of the extrapolation
metric parameters.
\section{Determining extrapolation box and constructing boundary conditions}
For the entire series of calculations we use the same rectangular 3D box 
$[214.56\times 213.12\times 183.6]$Mm. The center of its bottom face corresponds to approximate center of AR 11158
at
$-20.85^o$ latitude and  $34.9^o$ longitude. The $X$ axis of the bottom face is directed towards the North Pole.
Coordinates $x$, $y$
correspond to the Lambert equal area projection  \citep{Calabretta,Thompson}. The initial  $z$ -coordinate
corresponds to the solar radius. For calculation, grid  $(299\times297\times256)$is used, with a relevant pixel
size of $720$ km.
For extrapolation, we use  SDO/HMI (hmi.B\_720s) vector magnetograms without azimuthal ambiguity.
We remap data of the field of view magnetograms ($B_x,B_y,B_z$) to the box coordinate system on the bottom
boundary of the box.  
Thus, we will consider BVs  for the extrapolation problem at the bottom boundary as given.
To specify BVs at the other boundaries of the box, we use a potential solenoidal field, whose normal component
coincides at the bottom boundary with the magnetic field normal component defined above.  We calculate the
potential field with the Fourier decomposition algorithm similar to \citep{Alissandrakis} using FFT. At the same
time, in order avoid the zero harmonic problem in case the initial data give non-zero total flow of the normal
field, we specify the normal component on the extended region that contains 4 initial quadrants of the bottom
boundary of our box. In the first lower quadrant we set boundary values of our box. In the diagonal one – the same
boundary values.
In the rest quadrants there are boundary values of opposite sign. Next, we calculate the potential field in 3D box
only above the first quadrant. Thus, we ensure that the calculated normal component coincides with the one
specified at the bottom boundary. The potential field calculated in this way will be referred to as the field of
$BVP_{potI}$

Using the   $BVP_{potI}$ field, we define boundary values for  ($B_x,B_y,B_z$) at the other boundaries of the box.
We use the same potential field to define the starting field in 3D box for OPTI extrapolation exactly according to
the algorithm described in  \cite{Wheatland} without using the weight function.  To calculate the OPTI field free
energy (before postprocessing), further we also use this potential field. We calculate energies of OPTI and 
$BVP_{potI}$ fields in the root box 
$[169.26\times167.766\times159.12]$ Mm with grid $(235\times233\times221)$. 

To arrange Postprocessing II and to calculate free energies of Postprocessing I and Postprocessing II fields, we
need a potential field corresponding to the given normal field on all faces of the box
($BVP_{potII}$). To obtain this potential field, we use the code described in \cite{box}.

Further, for convenience, we will use the following abbreviations:
\begin{itemize}
\item $B_{all}$ - full boundary of the box.
\item $B_b$ -bottom boundary of the box. 
\item  $B_{st}$ - boundary including side and upper boundaries of the box. 
\item $BV_{all}$ - boundary values  $B_x,B_y,B_z$ at $B_{all}$.
\item $BV_b$ - boundary values $B_x,B_y,B_z$ at $B_b$.
\item  $BV_{st}$ - boundary values $B_x,B_y,B_z$ at $B_{st}$.
\item $BVn_{all}$ - boundary values  $B_n$ at $B_{all}$.
\item $BVn_b$ - boundary values $B_n$ at $B_b$.
\item  $BVn_{st}$ - boundary values $B_n$ at $B_{st}$.
\item $BVt_{all}$ - boundary values of field components  ($B_t$) at $B_{all}$.
\item $BVt_b$ - boundary values $B_t$ at $B_b$.
\item  $BVt_{st}$ - boundary values $B_t$ at $B_{st}$.
\end{itemize}
    Therefore, the OPTI extrapolation problem is formulated as NLFF extrapolation with $BVP_{OPTI}$: when $BV_b$
    from vector magnetograms data and \\  
$B_{st}$=$B_{st}(BVP_{potI})$.
\section{Features of  OPTI extrapolation code}
The code we use for OPTI extrapolation corresponds exactly to the  \cite{Wheatland} algorithm without using the
weight function with the fixed shift of time parameter of the optimization process 
$dt=0.05\Delta^2$, where $\Delta$ is the pixel size of grid.The number of optimization process steps is always
constant and amounts $400000$. It is with this number of steps that the required quality of extrapolation results
is achieved. 
The algorithm is built on parallelized calculations. Calculation of one problem was performed using a
supercomputer with 36 processors. One extrapolation for a box with the characteristics described above is
performed within $\sim20$ h. 

Since our extrapolation (R1) results can be compared with those (R2) by \cite{Sun}), due to their significant differences, we will list common and different elements of R1 and R2:
\begin{itemize}
\item for R1 and R2, extrapolation boxes and grids are almost the same. 
\item  in R1 and R2, the potential field is calculated in different manner (their energies are close to each
other).    
\item preprocessing is not used in R1, but it is used in R2 (not to change measurement data is our principled
position). 
\item in R1, weight function is not used in the  \cite{Wheatland} minimizing functional, in R2 – it is used. 
\item R1 uses a fixed $dt$ step and a fixed number of steps, R2 uses automatic adjustment of $dt$ step and the end
of the optimization process based on criterion of convergence of the functional $L$ metric from \cite{Wheatland}
. In the second case, apparently, the optimization ends too soon, bumping into the local minimum of $L$ in the
functional space. In the first case, the optimization process occurs with regular reduction of $L$, without the
requirement of reducing it at each step. Most likely, this last point leads to a significant difference in OPTI
extrapolation.
\end{itemize}
\section{Postprocessing}
Suppose we have a non-solenoidal OPTI magnetic field $\bf{B}$ complying with  \\$BVP_{OPTI}$. Since the field flow
is zeroth, the following equations are valid at  $B_{all}$:

\begin{equation}\label{eq1}
\int_S{\bf B}\cdot d{\bf s}=\int_V\nabla \cdot {\bf B}dv=0.
\end{equation}
For this   $BVP_{OPTI}$ field, we will consider two ways of cleaning the divergence.
\subsection{Postprocessing I}
According to Formula (B.4) from \cite{Valory_2013}, from $\bf{B}$ we can distinguish the following solenoidal
${\bf B}_{sI}$ component:
\begin{equation}\label{eq2}
{\bf B}_{sI}={\bf B}+\widehat{{\bf z}}\times\int_z^{z_2}(\nabla \cdot {\bf B})dz',
\end{equation}
where  $z_2$ is coordinate  $z$ of the box upper boundary.  This equation  (\ref{eq2}) does not change components
$B_x,B_y$ throughout the box.  If field   $\bf{B}$ were strictly solenoidal, the error in $B_z$ would have the
order of magnitude
$B_z$ at the top boundary of the box, i.e. it would be negligible (in numerical implementation of Equation
(\ref{eq2})). Non-solenoidality of  $\bf{B}$ leads to a significant change in $B_z(z=z_0)$, therefore, the
reference potential field ($BVP_{potII}$) for ${\bf B}_{sI}$ should change significantly.

\subsection{Postprocessing II}
Postprocessing II is to find the solenoidal field of  $\bf{B}_{sII}$ in the form
\begin{equation}\label{eq3}
{\bf B}_{sII}={\bf B}_{sI}+\nabla \psi',
\end{equation}
where the solenoidal potential field $\nabla \psi'$ provides equality 
\begin{equation}\label{eq3}
({\bf B}_{sII})_z=B_z \qquad  for \quad  z=z_0,
\end{equation}
i.e. $\nabla \psi'$ provides equal  $BV_b$ of fields  $\bf{B}$ and ${\bf B}_{sII}$ compensating for  $BV_b$
changes in the  ${\bf B}_{sI}$ field. We find field  $\nabla \psi'$ by solving the problem of  $BVP_{potI}$
numerically.
Thus, in Postprocessing II, situation is opposite – at the bottom boundary of the box, the transverse field
changes significantly while the box normal component remains unchanged. Changes in  $BV_{st}$ are usually
negligible for a compact magnetic region far enough from the boundaries. In addition, we don't in urgent need of
preserving  $BV_{st}$ that have rather artificial nature of origin. In practice, energies of potential reference
fields of $\bf{B}$ and ${\bf B}_{sII}$ are very close to each other.

We should note a very important point of the presence of the common feature for procedures of Postprocessing I and
II, to which we will return again a little later – both procedures accurately retain the observed $z$ component of
current  $\bf{J}$ on $V_b$. 
\section{Analysis of OPTI results and their modifications}
\subsection{Configuration of field lines}
\begin{figure}
\includegraphics[width=0.48\textwidth]{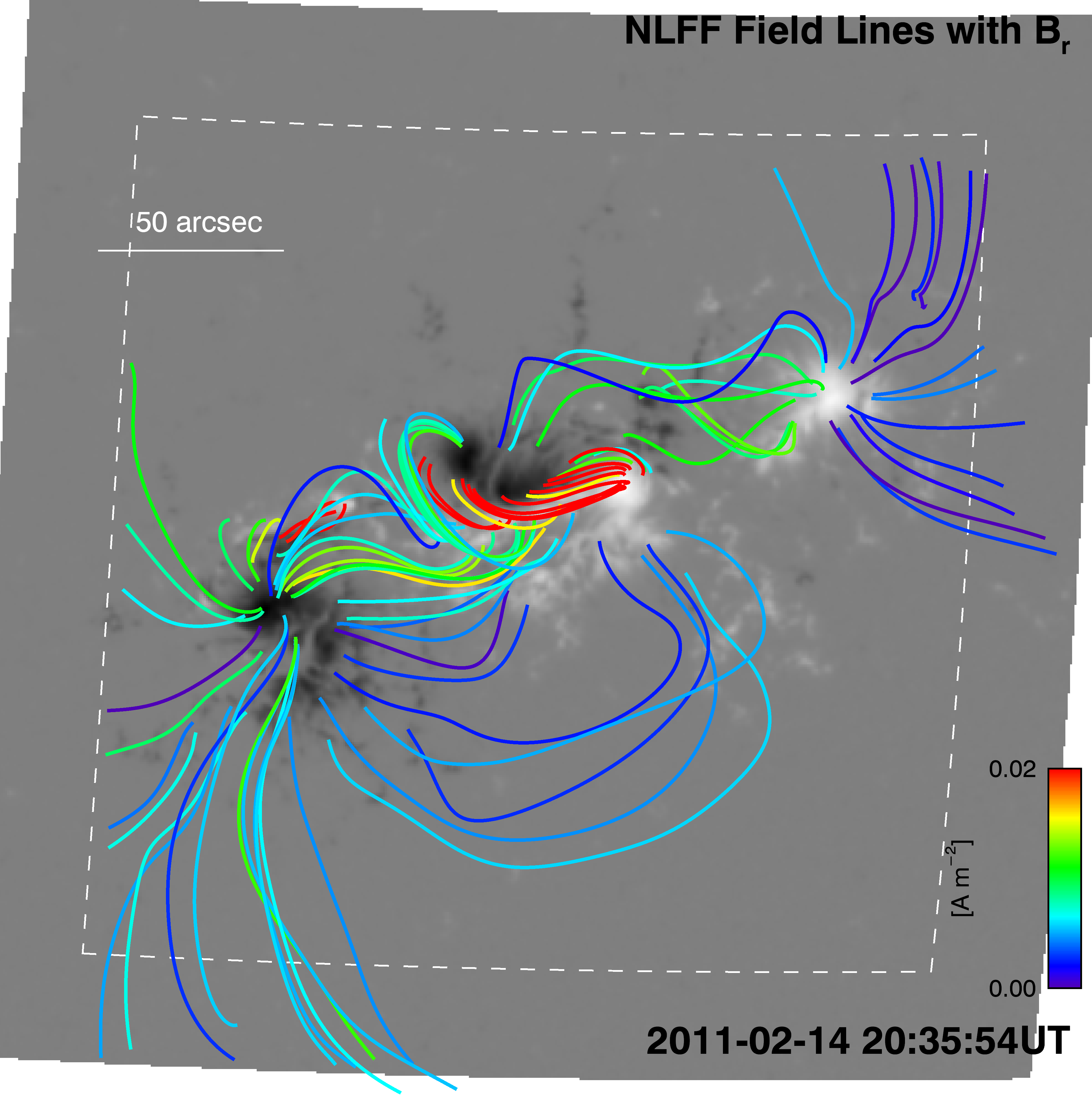}
\includegraphics[width=0.48\textwidth]{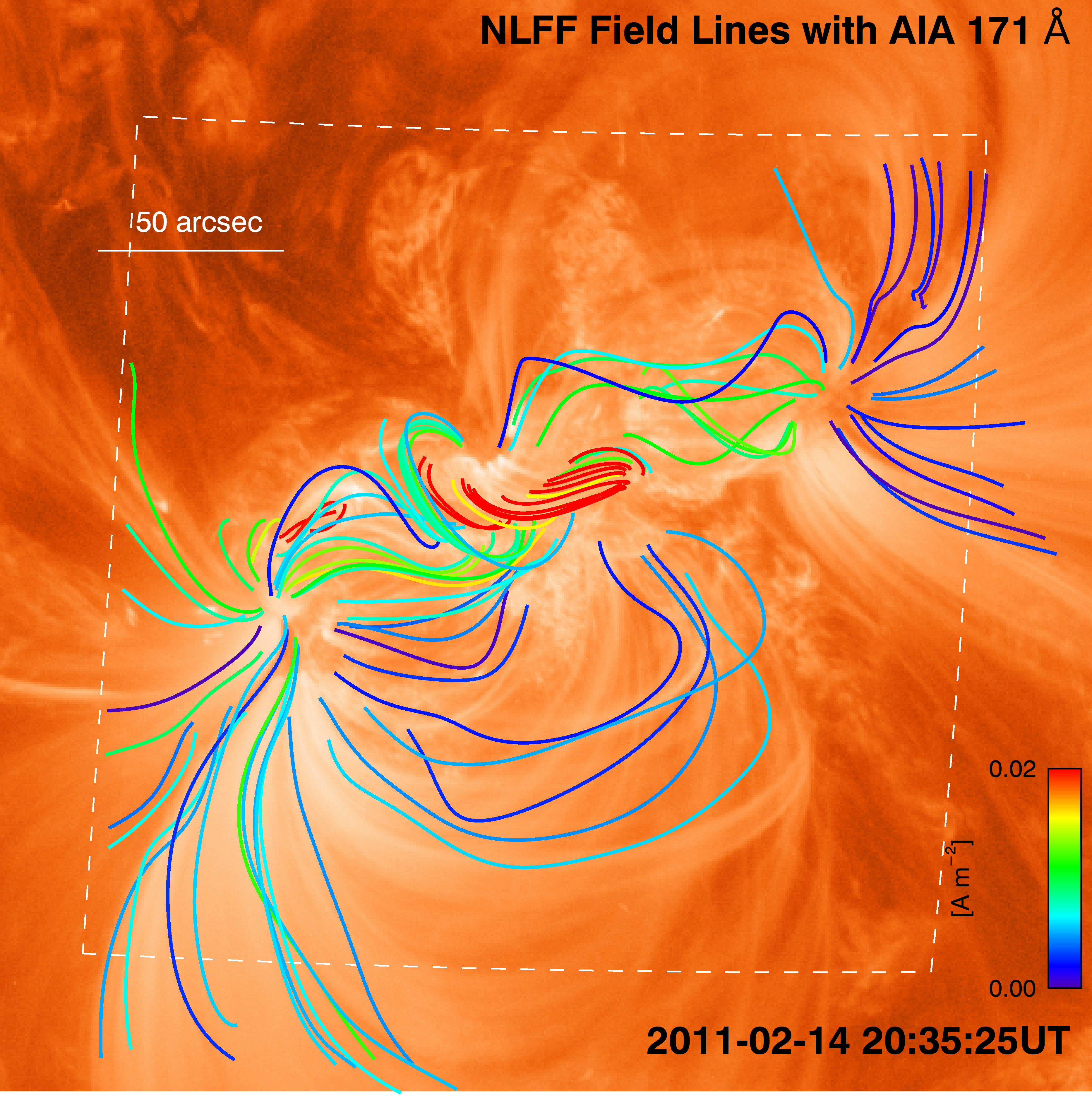}
\caption{ A case without postprocessing. Left: Selected field lines from the NLFFF extrapolation plotted over a
frame of Brmap. The lines are color-coded by vertical current density at their footpoints (see the color bar).
Dashed line outlines the bottom of the core domain of NLFFF extrapolation. Right: Same field lines over the AIA
image frame.}
\label{fig1}
\end{figure}

\begin{figure}
\includegraphics[width=0.48\textwidth]{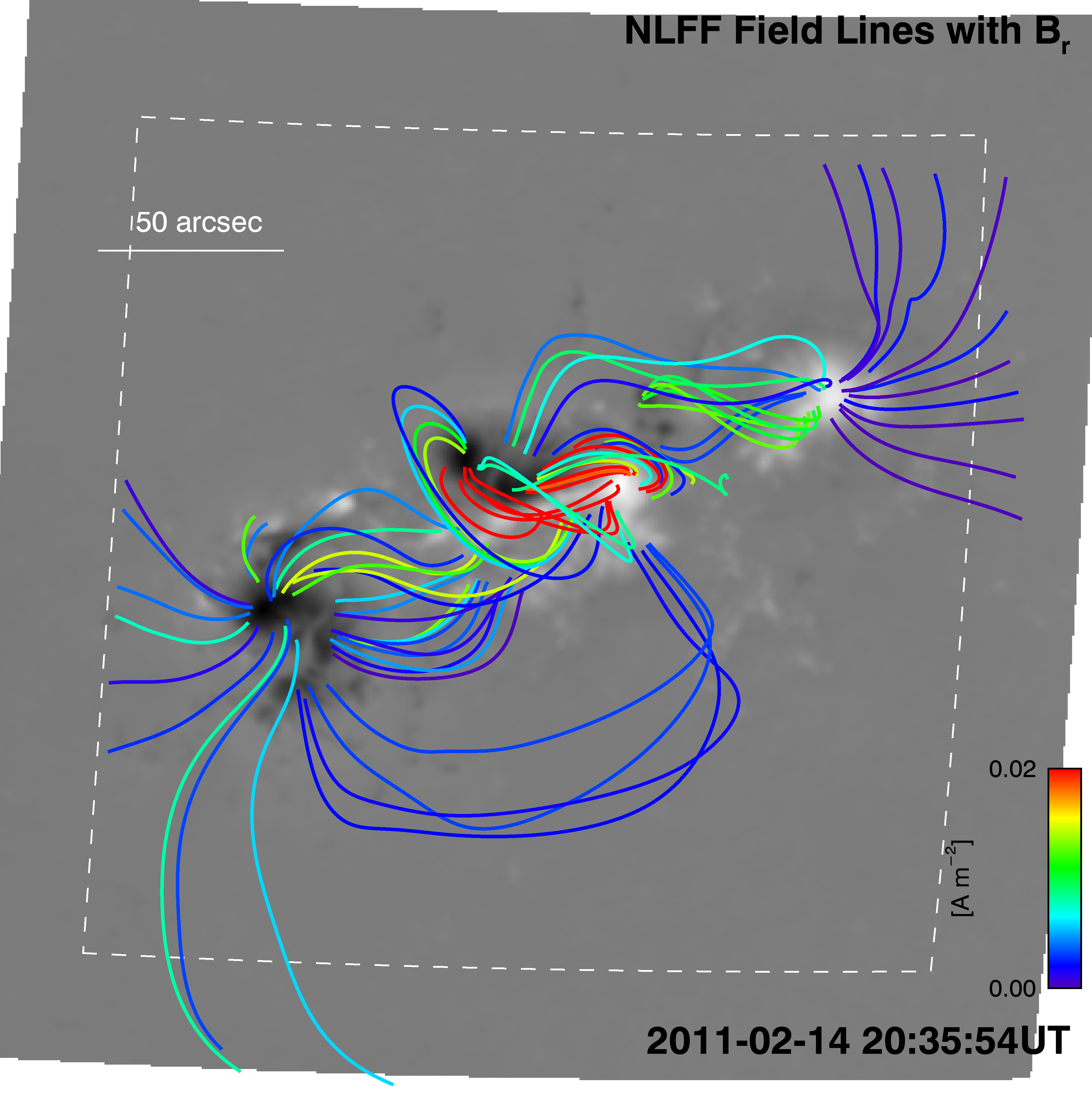}
\includegraphics[width=0.48\textwidth]{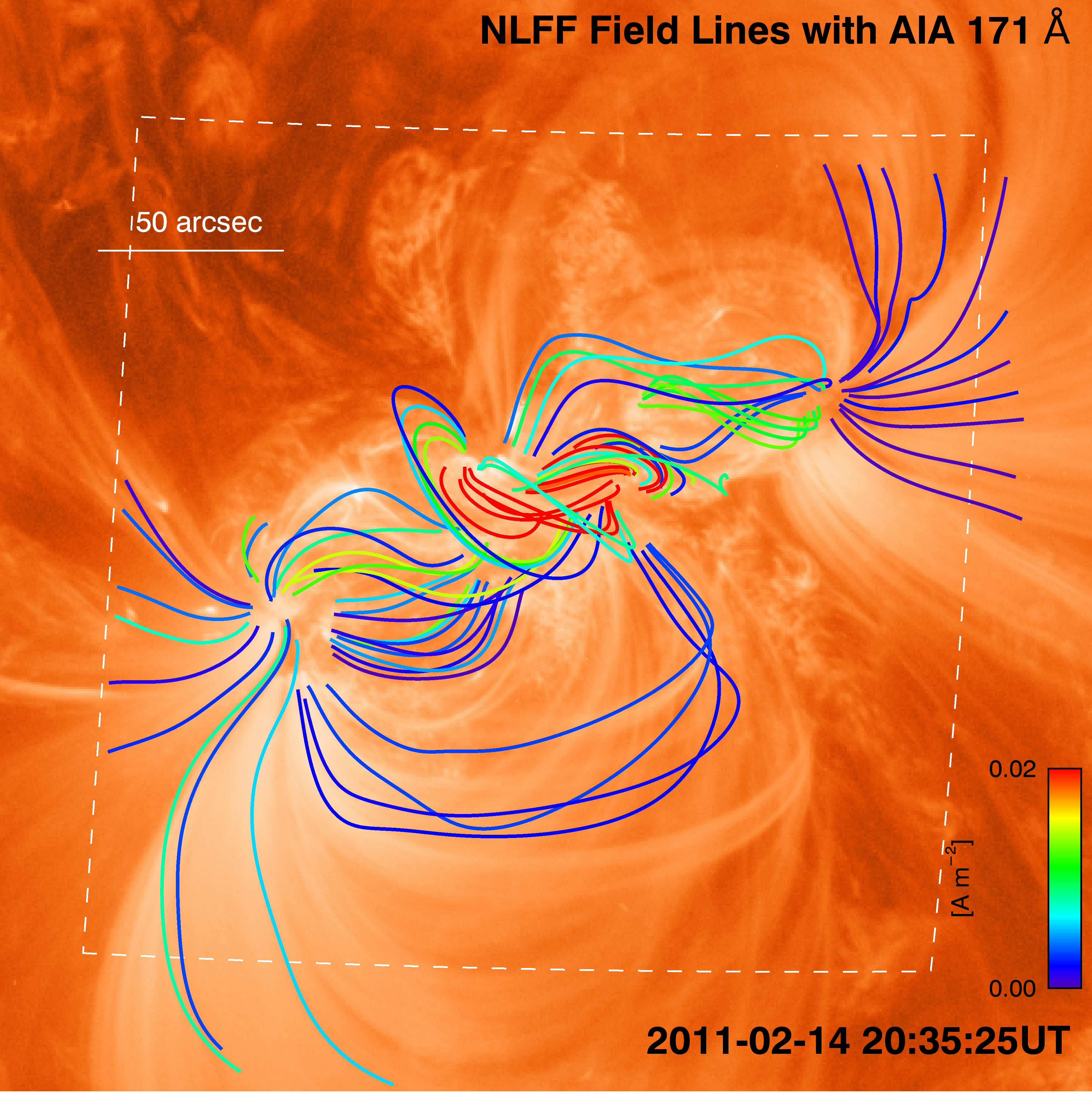}
\caption{Same as in Figure \ref{fig1} for the Postprocessing I case.} 
\label{fig2}
\end{figure}

\begin{figure}
\includegraphics[width=0.48\textwidth]{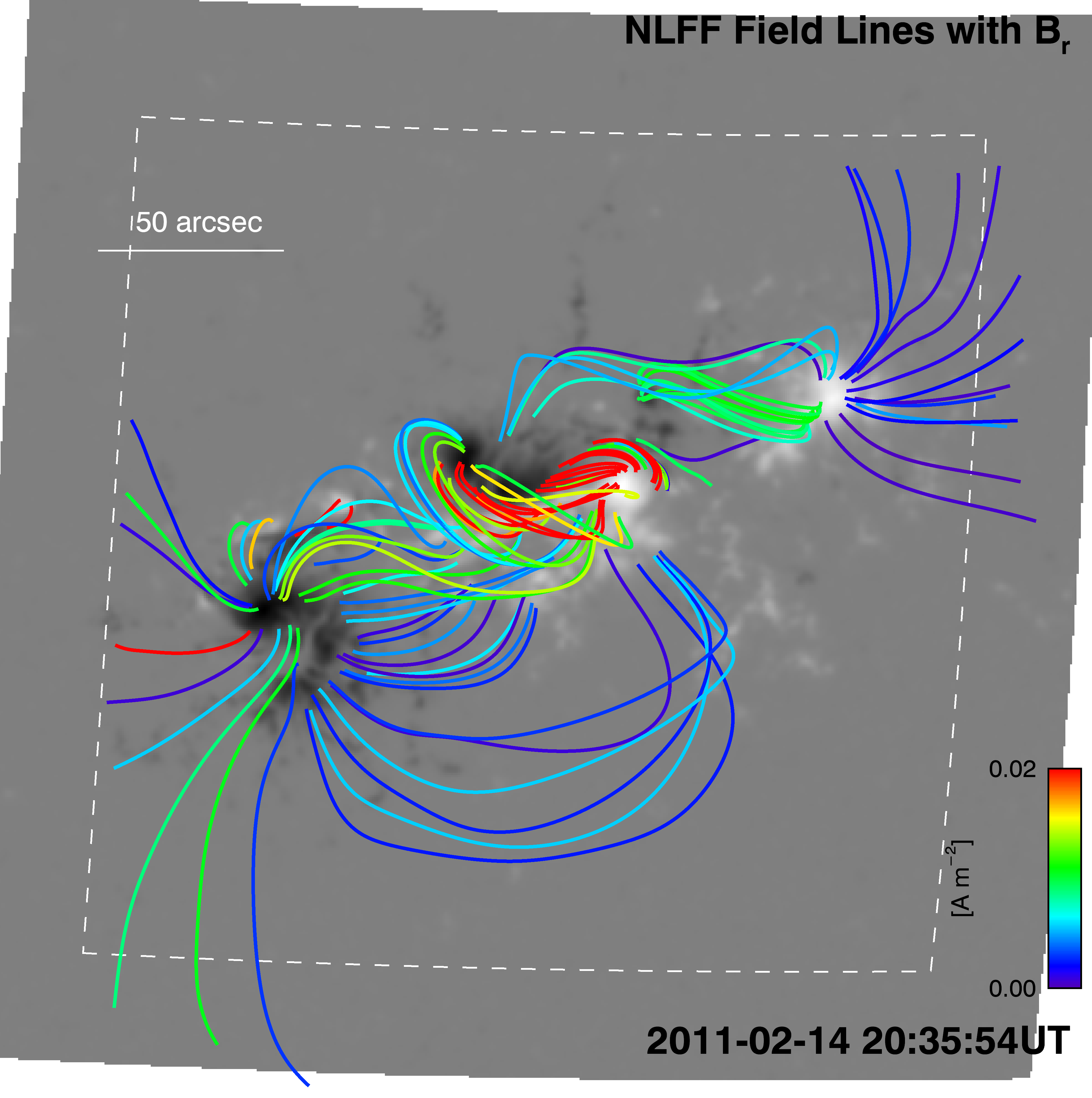}
\includegraphics[width=0.48\textwidth]{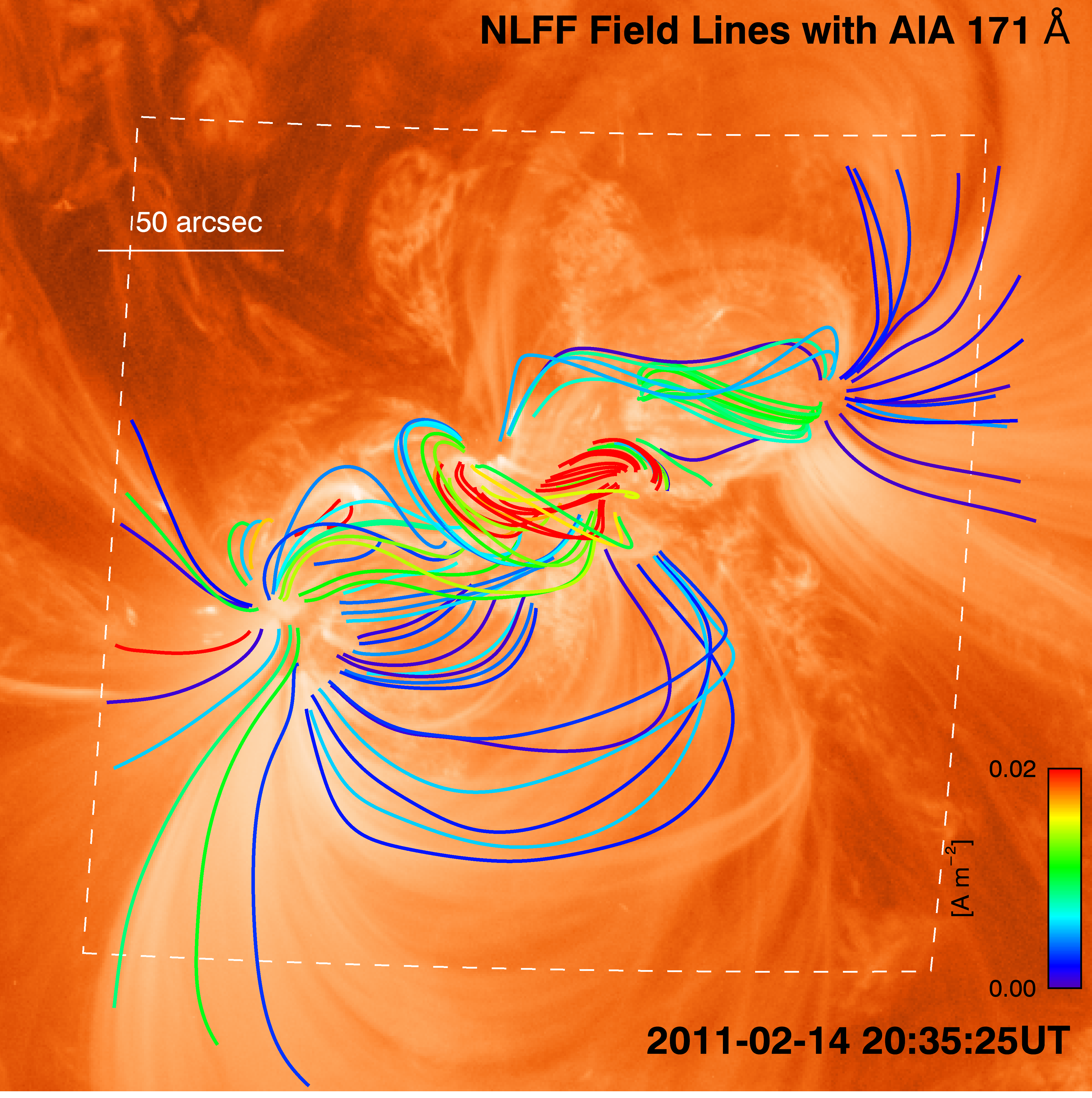}
\caption{Same as in Figure \ref{fig1} for the Postprocessing II case.} 
\label{fig3}
\end{figure}

For the same time point, for which the overall mapping of the field lines configuration for AR 11158 was given in
\cite{Sun}, we give a similar representation of field lines in Figures   \ref{fig1}-\ref{fig3}.We see that our
line images are well correlated with those in   \cite{Sun} and result in approximately the same compliance with
the real structure of field lines highlighted on the AIA image. It is quite difficult to prefer one of our results
or the result by \cite{Sun}. Note that without postprocessing, OPTI field results in many broken non-real field
lines (mostly coming from the weak field regions).  However, we see that the basic structure is shown quite well.
Procedures of Postprocessing I and II yield “smoother” fields. The expected lines are easier to find. From value
judgment, Postprocessing II results in a slightly “smoother” field in relation to field of Postprocessing I.
\subsection{Metric analysis} 
We analyze the metrics of calculation results for the same point in time as in the previous subsection  20011-02-14 20:36 UT. We use the following set of metrics for analysis.
\begin{equation}\label{eq5}
E=\frac{1}{8\pi}\int_V{\bf B}^2dv,
\end{equation}
where $E$ is the energy field.
\begin{equation}\label{eq6}
E_{pot}=\frac{1}{8\pi}\int_V{\bf B}_{pot}^2dv,
\end{equation}
where  $E_{pot}$ is the energy reference potential field calculated as  $BVP_{potII}$ for relevant  ${\bf B}$
\begin{equation}\label{eq7}
E_{free}=E-E_{pot},
\end{equation}
where $E_{free}$ is the value corresponding to determination of free energy for solenoidal field  {\bf B}.
\begin{equation}\label{eq8}
E_{free}^*=\frac{1}{8\pi}\int_V\left({\bf B}-{\bf B}_{pot}\right)^2dv,
\end{equation}
where $E_{free}^*$ matches  $E_{free}$ from Equation (\ref{eq7}) for solenoidal field  {\bf B}.
\begin{equation}\label{eq9}
\varepsilon_1= \left\vert \frac{E_{free}-E_{free}^*}{E}\right\vert, 
\end{equation}
where $\varepsilon_1$ describes the relative contribution of non-solenoidality in  {\bf B}.  
\begin{equation}\label{eq10}
\varepsilon_2= \left\vert \frac{E_{free}-E_{free}^*}{E_{free}}\right\vert. 
\end{equation}
Equation  (\ref{eq10}) exactly corresponds to definition of  $\varepsilon$ in Equation 15 from 
\cite{Mastrano_2018}. This value evaluates the error of the free energy related to field  {\bf B}
non-solenoidality. 
We also use
\begin{equation}\label{eq11}
\theta _{j}=\arcsin \left( \frac{\sum_i^N \left\vert {\bf J}\right\vert _{i}\sigma _{i}}{\sum_i^N \left\vert {\bf
J}\right\vert _{i}}\right) ,\quad \sigma _{i}=\frac{\left\vert {\bf J}\times {\bf B}\right\vert _{i}}{\left\vert
{\bf J}\right\vert _{i}\left\vert {\bf B}\right\vert_{i} },
\end{equation}

\begin{equation}\label{eq12}
f=\frac{1}{N}\sum_i^N\frac{\left\vert \nabla \cdot{\bf B}\right\vert_i}{6\left\vert
{\bf{B}}\right\vert_{i}}\Delta.
\end{equation}
The metrics of Equations (\ref{eq11}) and (\ref{eq12}) are similar to those introduced by  \cite{Wheatland}. The
former reflects deviation of  {\bf B} from the force-free approximation, the latter – solenoidality.

For the time investigated, metrics of Equations  (\ref{eq5})-(\ref{eq12}) are presented in Tables  (\ref{table1})
and (\ref{table2}).
\begin{table}
\caption{Metrics of energy characteristics }
 \label{table1}
\begin{tabular}{ccccc}
\hline
  & $E$ $(\times10^{32}erg)$ & $E_{pot}$ $(\times10^{32}erg)$& $E_{free}$ $(\times10^{32}erg)$& $E_{free}^*$
  $(\times10^{32}erg)$\\
 \hline
OPTI   &$10.45$&$8.46$&$1.99$&$3.54$  \\
+Postprocessing I   &$9.53$&$6.56$&$2.96$&$3.03$  \\
+Postprocessing II   &$11.41$&$8.49$&$2.92$&$3.03$  \\
 \hline
\end{tabular}
\end{table}
\begin{table}
\caption{Metrics of relative errors}
 \label{table2}
\begin{tabular}{ccccc}
\hline
  & $\varepsilon_1$ & $\varepsilon_2$& $\theta _{j}$ (Degree)& $f$\\
 \hline
OPTI   &$0.14$&$0.78$&$11.91$&$0.008$  \\
+Postprocessing I   &$0.0067$&$0.021$&$15.55$&$0.002$ \\
+Postprocessing II   &$0.0089$&$0.034$&$17.37$&$0.002$  \\
 \hline
\end{tabular}
\end{table}
We see that OPTI+Postprocessing I reduces, and \\OPTI+Postprocessing II, on the contrary, increases the energy
relative to OPTI by approximately the same values of  $10^{32} erg$. OPTI+Postprocessing II practically does not
change the potential field energy, while it decreases significantly in OPTI+Postprocessing I.  Free energies after
both types of postprocessing are equally (significantly) increased by the same order of magnitude $10^{32} erg$,
while the OPTI+Postprocessing II energy is  $~2\times10^{32} erg$ greater than the OPTI+Postprocessing I energy.

\begin{figure}
\includegraphics[width=0.48\textwidth]{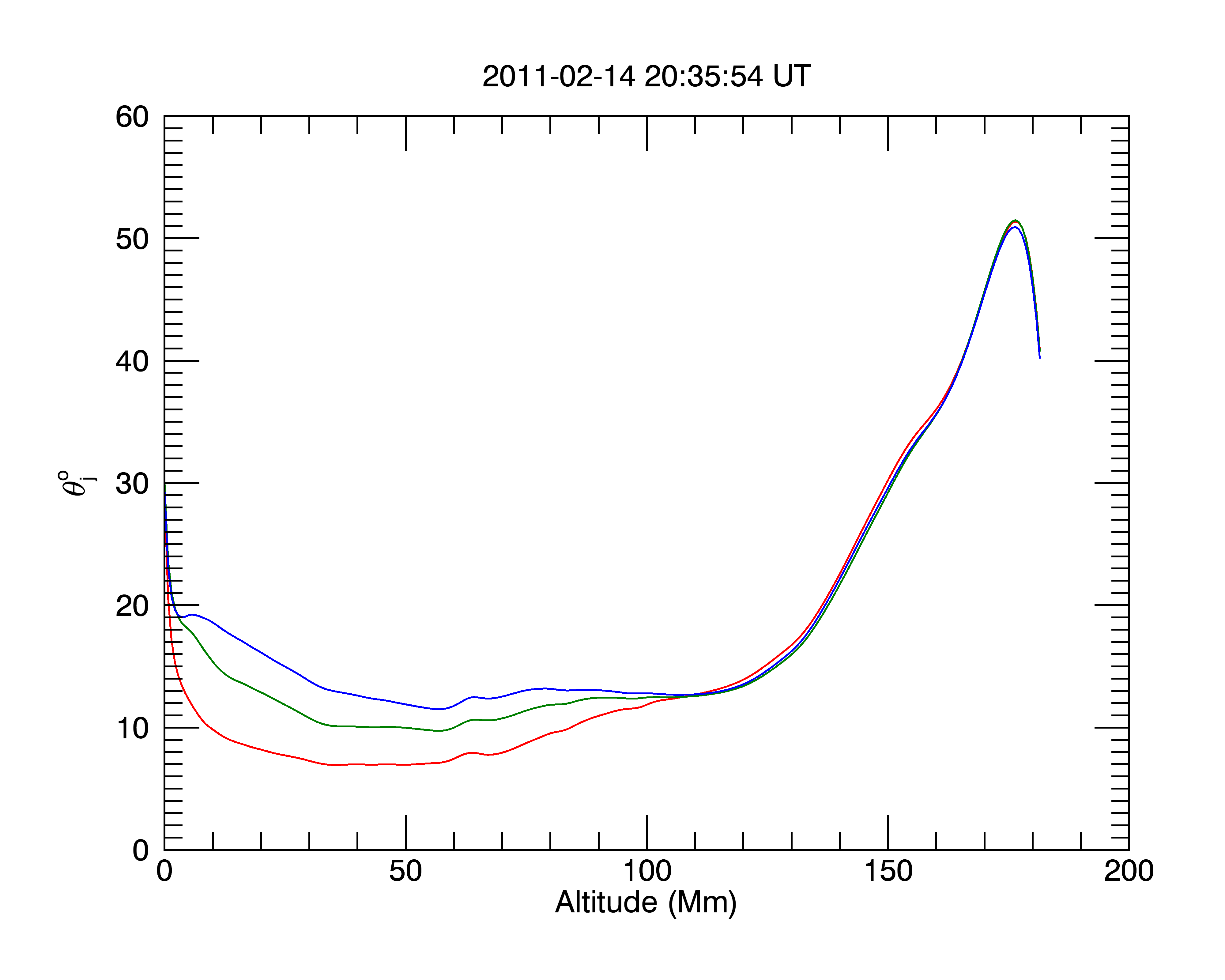}
\includegraphics[width=0.48\textwidth]{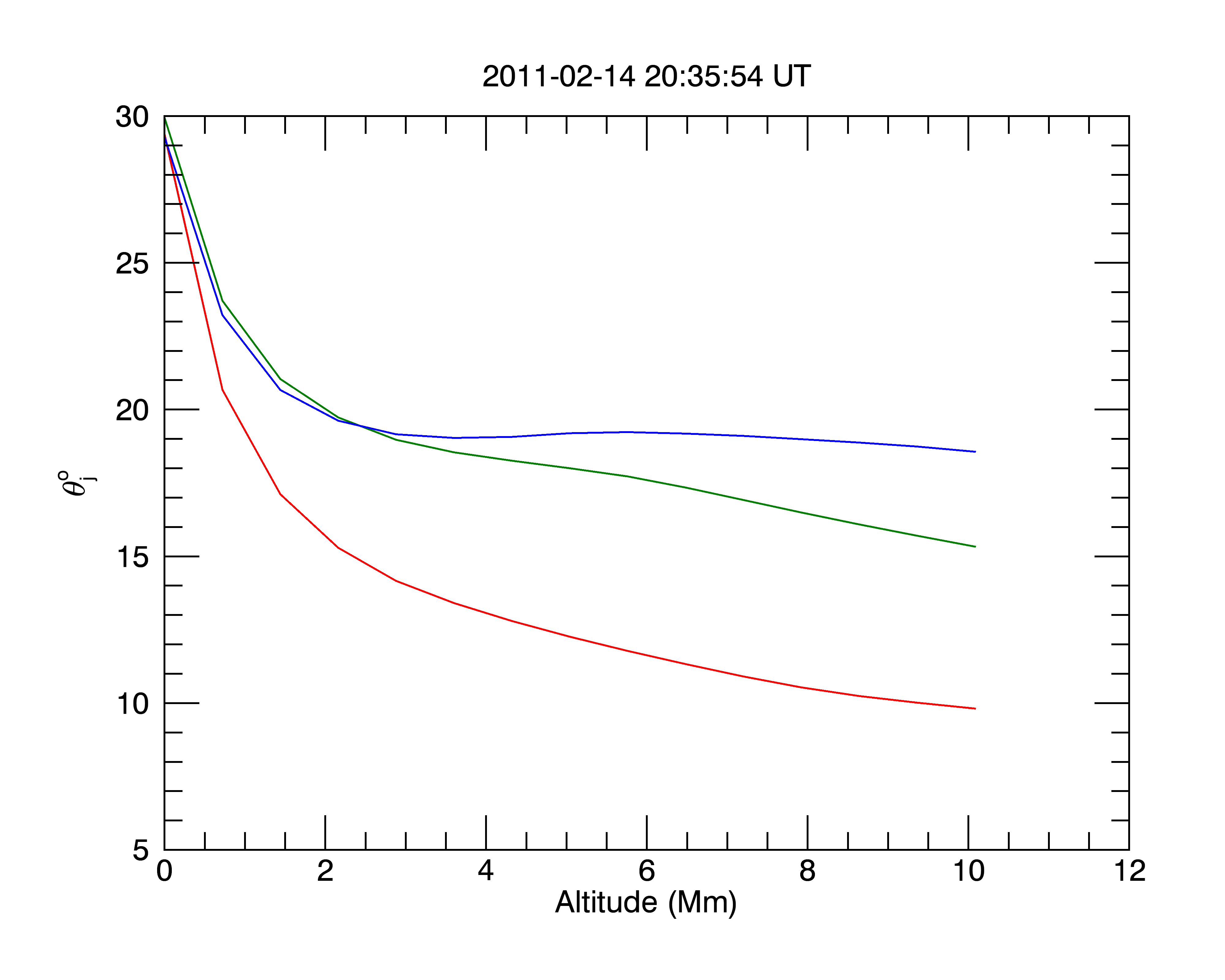}
\caption{Altitude dependence of  ${\theta _{j}}$: red - OPTI, green - OPTI+Postprocessing I, blue
-OPTI+Postprocessing II.} 
\label{fig4}
\end{figure}

The OPTI result gives a relatively small  $\varepsilon_1$, i.e. formally we can consider this solution close to
the solenoidal one. If this value were not small, notionally no compliance of OPTI extrapolation with the real
field could be expected.  At the same time, as we see from 
$\varepsilon_2$, we cannot estimate the real free energy with fair accuracy.  Note that 
$\varepsilon_2$ is consistent with the result obtained for OPTI in  \cite{Mastrano_2018}. Values   of
$\varepsilon_1$ and $\varepsilon_2$ for OPTI+Postprocessing I, II show that we succeed in getting rid of
non-solenoidality. Besides, values of  $\varepsilon_2$ are well consistent with similar ones 
\citep{Mastrano_2018} for NLFFF extrapolation methods operating with  the class of solenoidal fields. From $f$
metric (Table  \ref{table2}), we can also see the result of eliminating the non-solenoidality to the level of
numerical errors, we hope. After elimination of non-solenoidality we naturally obtain a certain increase in
${\theta _{j}}$ metric after postprocessing, indicating  deviation of the solution from the force-free
approximation. We cannot estimate, to which extent this value reflects the error or effect of the real deviation
of the field from the force-free field.  Applying Equation  \ref{eq11} to each  $z$ layer and displaying ${\theta
_{j}}$ in Figure \ref{fig4} , we provide better representation of the force-free characteristic of our results. 
We see that, starting at the same level, all three characteristics decrease with height. Postprocessing II shows
the highest value. Starting from the height of $\sim 110$ Mm, all three characteristics merge and grow to a large
value typical of the potential field. The OPTI algorithm may not significantly change the initial potential field
at these altitudes. In any case, the field at these altitudes is quite small and does not essentially contribute
to the energy we are interested in. 

\section{Evolution of AR 11158 energy in OPTI results}
\begin{figure}
\centerline{\includegraphics[width=0.7\textwidth]{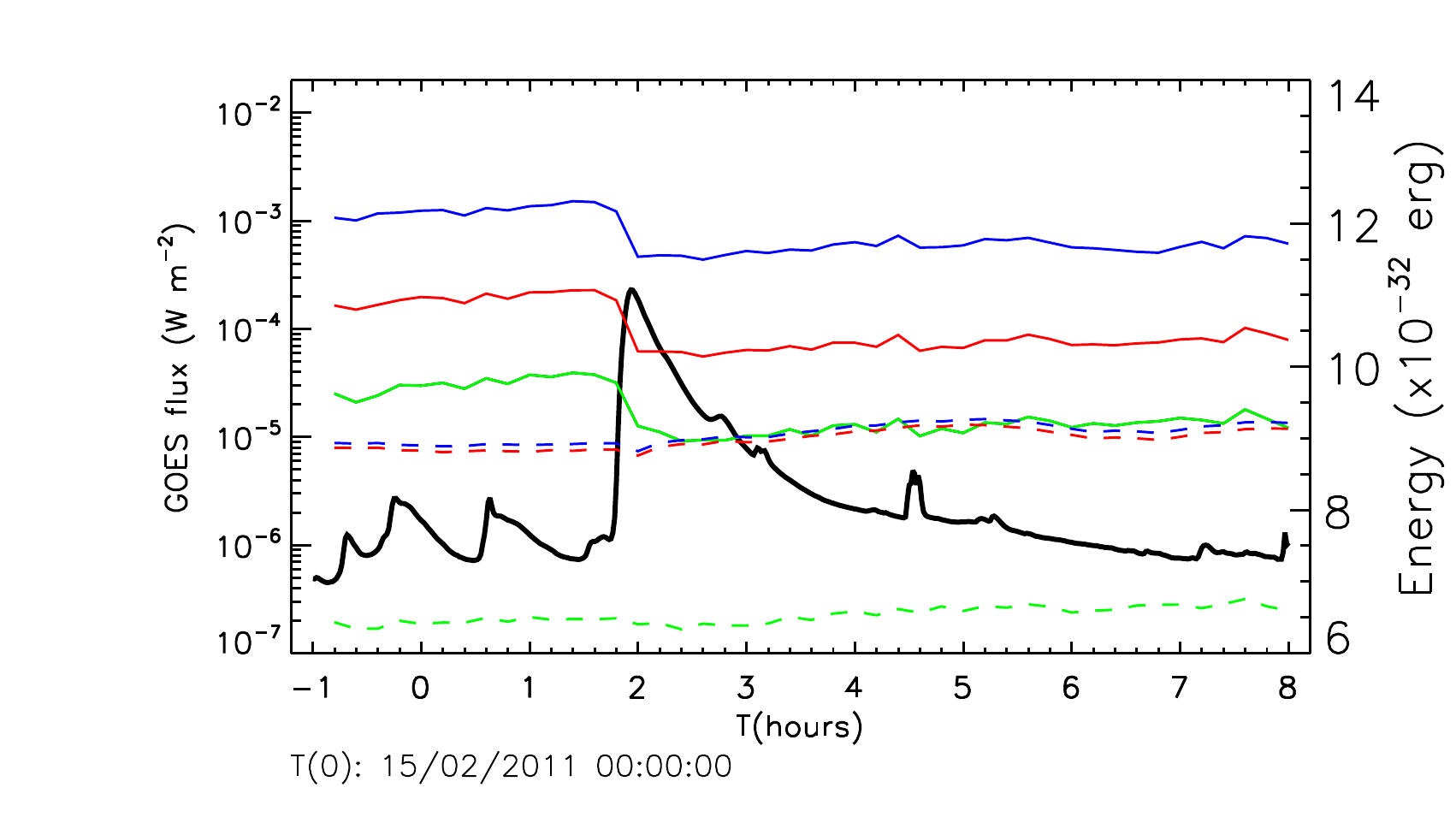}}
\caption{Time plot of energies for the AR 11158 X-class flare. Solid curve is the GOES flux profile. Solid colored
lines -$E$, dashed colored lines  - $E_pot$, red - OPTI, green - OPTI+Postprocessing I, blue -OPTI+Postprocessing
II.} 
\label{fig5}
\end{figure}

Consider the results of the NLFF extrapolation of AR 11158 in the time interval with the X-class flare that began
at 2011/02/15 01:44UT. Figure \ref{fig5} describes the time behavior of energies of all three of our
extrapolations (solid lines) and energies corresponding to the reference potential fields (dashed lines). These
dependencies reflect the X-flare event well, but they have significant   shifts along the energy axis, same as in
the case discussed in the previous section. The lowest energies result from OPTI+Postprocessing I, then OPTI and
the highest belongs to OPTI+Postprocessing II.  Energies of the reference potential fields of OPTI and
OPTI+Postprocessing II, as expected, are very close to each other.  Energies of the OPTI+Postprocessing I
reference potential field is lower by  $2.5\times10^{32}$ erg. For OPTI, energies of fields and reference
potential field were calculated (as it is normally done) in the root domain of the box, using $BVP_{potI}$ for the
OPTI reference potential field. For all other cases, energies were calculated for the entire box using
$BVP_{potII}$ for the reference potential fields
In general, we can say that energies calculated for the entire box or its root domain always have little
difference between each other, which can be clearly seen from graphs for the OPTI and OPTI+Postprocessing II
reference potential fields. They almost coincide with each other despite difference of their $BVn_{st}$.

\begin{figure}
\centerline{\includegraphics[width=0.7\textwidth]{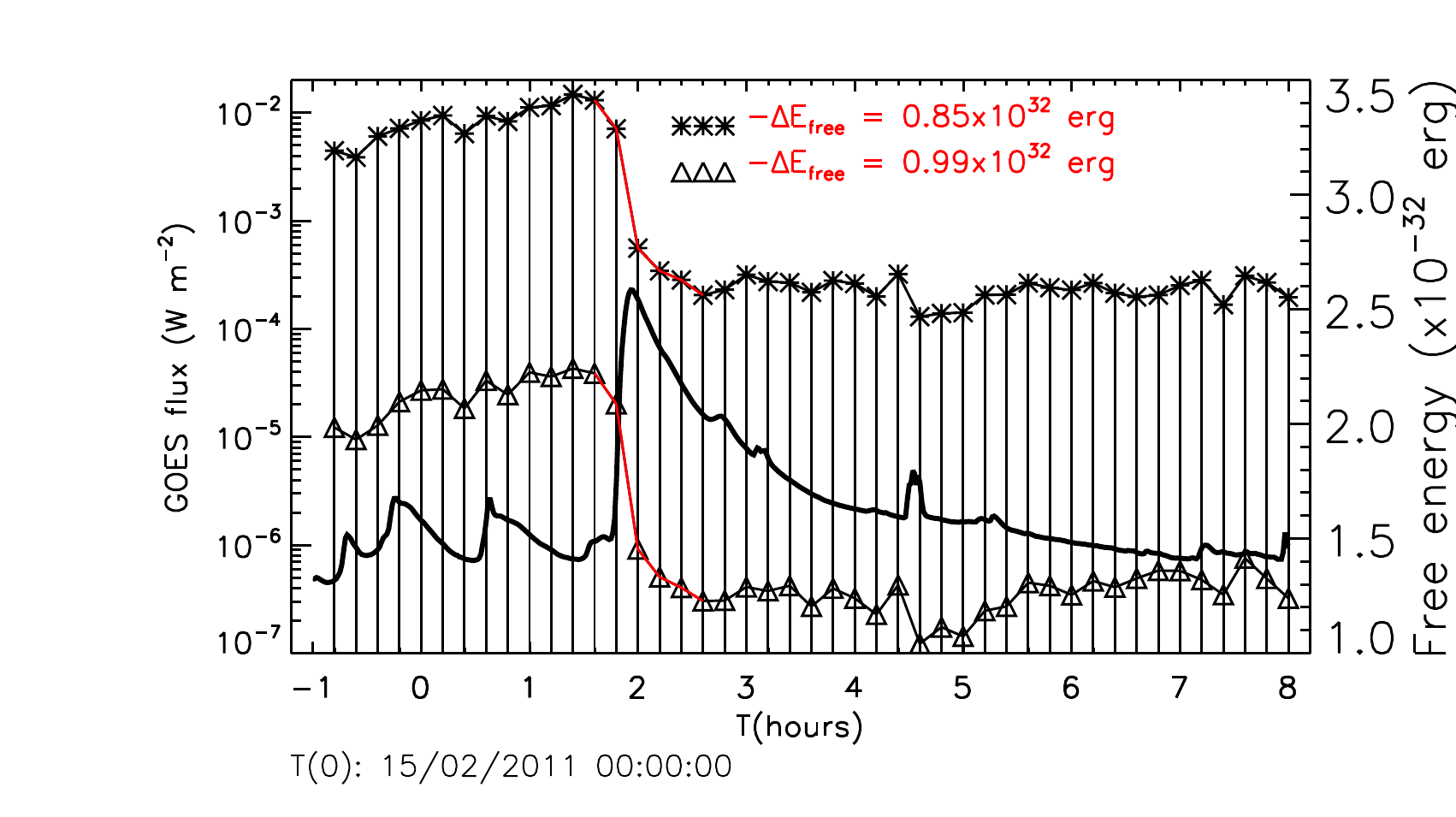}}
\caption{Time plot of free power for X-class flare of AR 11158. Solid curve is the GOES flux profile.Triangles –
for OPTI, asterisk – OPTI+Postprocessing I.} 
\label{fig6}
\end{figure}

\begin{figure}
\centerline{\includegraphics[width=0.7\textwidth]{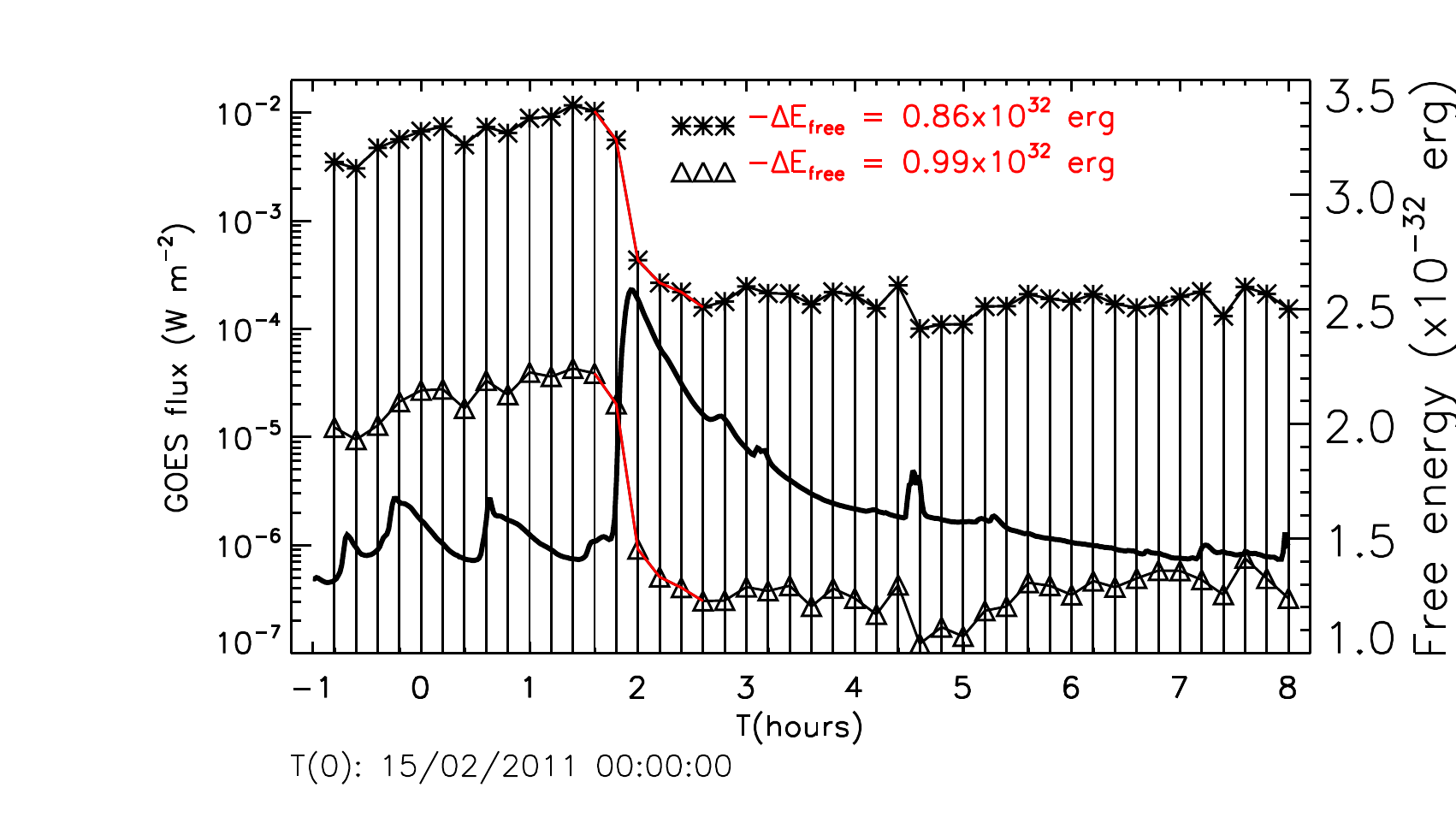}}
\caption{Time plot of free power for X-class flare of AR 11158. Solid curve is the GOES flux profile. Triangles –
for OPTI, asterisk – OPTI+Postprocessing II.} 
\label{fig7}
\end{figure}

\begin{figure}
\centerline{\includegraphics[width=0.7\textwidth]{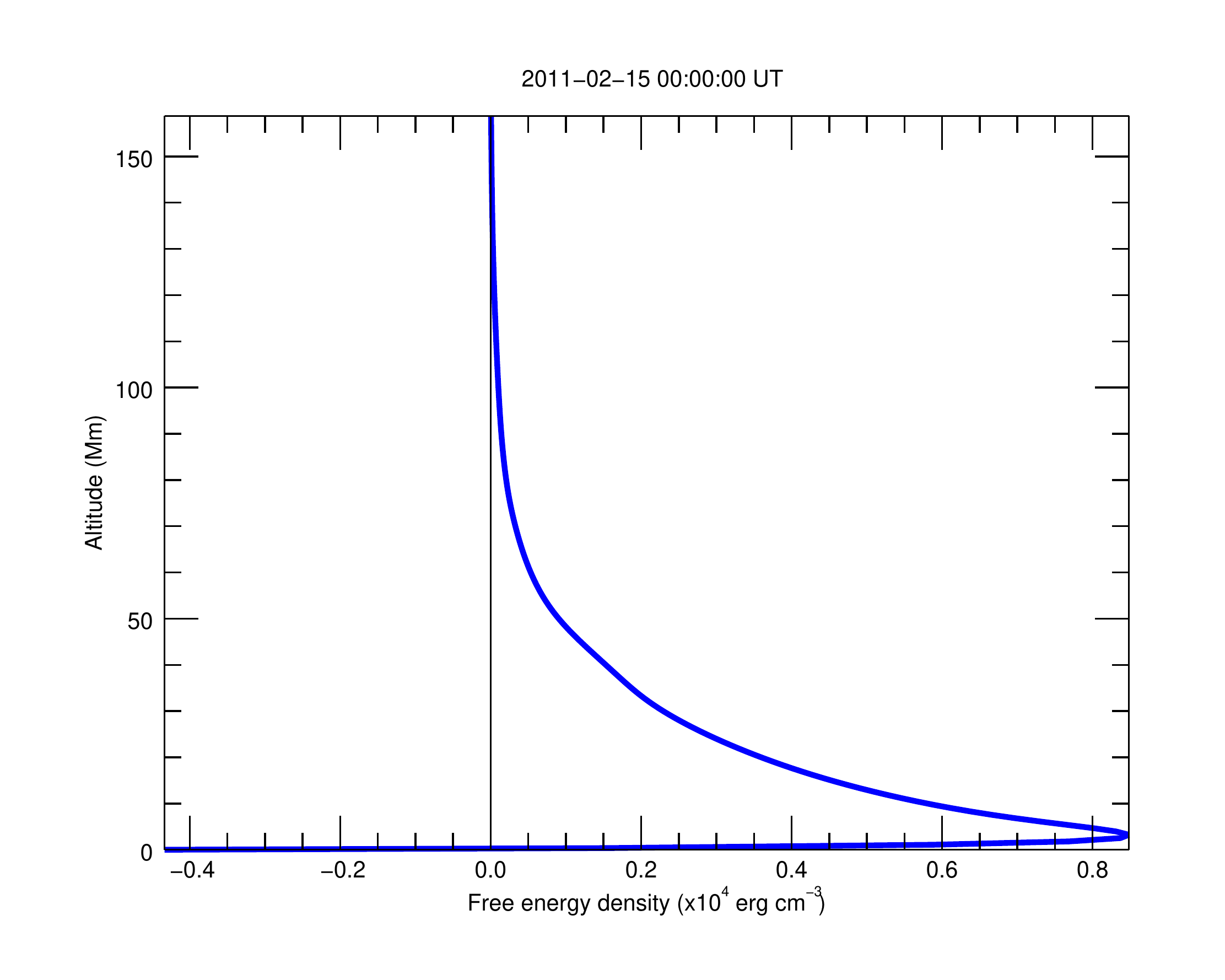}}
\caption{Height profile of OPTI mean free energy density of AR 11158 for T=2011-02-15 00:00:00 UT.}
\label{fig8}
\end{figure}

Figures  \ref{fig6} and \ref{fig7} describe the time behavior of free energies. Note that all three current
components of our solutions with zero $BVn_{all}$ differ from each other by only some potential (for OPTI
non-solenoidal) fields. Therefore, due to the solenoidality of OPTI+Postprocessings I, II, their free energies
must coincide with each other.  That's what we can observe in Figures \ref{fig6} and \ref{fig7}. Influence of
non-solenoidality in OPTI leads to a general decrease in free energy without fundamental changes in general nature
of its behavior.
As can be seen from Figures  \ref{fig6} and \ref{fig7}, all three dependencies of free energy give the order of
its dropping
$-\Delta E_{free}~10^{32}$ erg during one hour after the X-flare onset.  This corresponds to the estimate of
released energy in X-class events \citep{Hudson,Bleybel}. To explain significant difference between this result
and that in
\cite{Sun}, we compare the height dependence of the mean free energy density in Figure  \ref{fig8} with the same
dependence for the same point in time shown in Figure 4 from \cite{Sun}. While heights and the value maxima
coincide, our mean free energy density has a much larger scale of descent.  In \cite{Sun}, this value shows the
field proximity to the potential field starting already at $~20Mm$ heights. We have the same situation developing
much higher – at   $\sim 100$Mm altitudes (which is also consistent with the behavior of the 
$\theta _{j}$ force-free parameter in Figure \ref{fig4}).    This difference in behavior of the mean free energy
density seems to be related, as mentioned in Section 3, to the earlier termination of the optimization process in
\cite{Sun}.
In our case, we go further, thus calculating free energy more accurately.

\begin{figure}
\centerline{\includegraphics[width=0.7\textwidth]{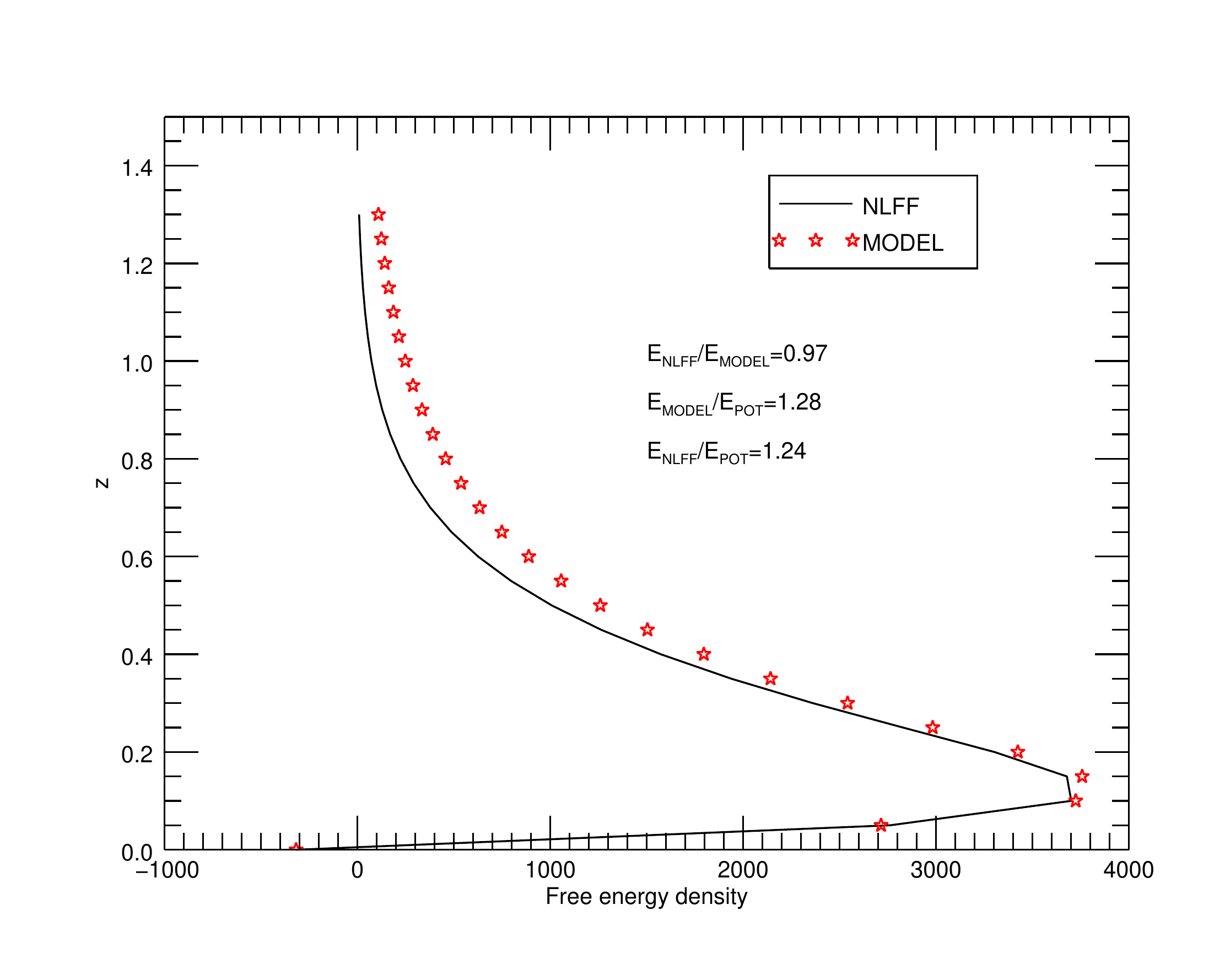}}
\caption{Height profile of the mean free energy density for the Low model.}
\label{fig9}
\end{figure}

Another important difference from  \cite{Sun} is manifested in negative values we obtain (for the entire time
interval in question) of the mean free energy density at zero height ( Figure  \ref{fig8})).
This result is interesting from the physical point of view. As it was shown in 
\cite{Livshits} ), for any force-free field above the sphere the following consequence of the virial theorem is
true – on the sphere  ($r=R$)), a field is more radial in relation to the reference potential field:
\begin{equation}\label{eq13}
\int_{S}\left( {\bf B}_{pot}\right) _{t}^{2}ds>\int_{S}({\bf B}_{t})^{2}ds, \qquad (r=R). 
\end{equation}
From Equation \ref{eq13}, considering equality of the normal components  $({\bf B})_n$ and $({\bf B}_{pot})_n$ it
automatically follows:
\begin{equation}\label{eq14}
\int_{S}({\bf B}^2-{\bf B_{pot}} ^2)ds <0, \qquad (r=R).
\end{equation}
Since the mean free energy density is proportional to the integral of Equation  \ref{eq14} , it should also be
strictly negative for the force-free field above the sphere. In the case of rectangular box geometry, the
statement above, strictly speaking, is not necessarily valid.  However, the real bottom boundary of our box
corresponds to the sphere, magnetic regions are mostly compact, and generally strong fields should be force-free
on the photosphere. Hence, negativity of the mean free energy density at zero altitude is likely to manifest
itself.  That's what we have.   Interestingly, the  \cite{Low} model has the same result for rectangular box
geometry  $[61\times 61\times 31]$, $[36\times36\times1.5]$Mm,  
with parameters   $l=0.5$, $\varphi
=1.4$, core ($[46,46,27]$,$[2.256\times2.256\times1.3]$Mm) and $\max(B_z)=800 Gs$ (see  Figure \ref{fig9}). 
In Figure \ref{fig9}, we see that this property of mean free energy density to take negative values near zero
altitude is equally manifested not only in the numerical OPTI result, but also in the analytical field model.  

\section{Conclusion}
Evaluation of the OPTI extrapolation validity has shown significant influence of partial non-solenoidality on the
free energy OPTI results. This effect is displayed in a significant shift in the free energy, but does not
significantly affect its time behavior character. Both with and without divergence cleaning, the OPTI
extrapolation gives the same  reasonable estimates of changes in the active region energy during powerful eruptive
X-class events. The divergence cleaning of OPTI results, regardless of the implementation method, leads to the
same values of free energy, only the energy levels being changed. Thus, we can trust behavior of changes in free
energy, both with and without divergence cleaning. Converting  the non-solenoidal part into the force component,
divergence cleaning is beneficial to  the obtained results. It significantly reduces the number of non-realistic
entangled magnetic lines, without essential changes in the field structure. Divergence cleaning (Postprocessings
II), which preserves the photospheric field normal component, does not change the reference potential field
energy. Divergence cleaning (Postprocessings I), which preserves the photospheric field transverse components,
reduces significantly the energy of its reference potential field. In terms of undesirability of changes in the
photospheric field normal component and obtaining high energies of the potential field and energies of the
calculated field, Postprocessing II looks more preferable. 
In our opinion, the main advantage of the OPTI approach to NLFF extrapolation (both with and without divergence
cleaning) is preserving the unchanged photospheric $({\bf J})_z$ current component, which, in fact, allows  to
obtain adequate estimate of the real energy release of the active region. From this point of view, a significant
difference can be expected between the reality and results of NLFF extrapolations allowing modification of the
$({\bf J})_z$ component. This, for example, takes place in the approach based on the algorithms of Grad Rubin
class. For the same reason, preprocessing (the procedure removes most of the net force and torque from the data,
\cite{Wiegelmann_2006}) often used in practice (among all, to OPTI extrapolation) is undesirable, because it 
inevitably leads to local changes in  $({\bf J})_z$. 
\begin{acks}
The work was performed with budgetary funding of Basic Research program II.16.The authors thank Irkutsk
Supercomputer Center of SB RAS for providing the access to HPC-cluster
Akademik V.M. Matrosov (Irkutsk Supercomputer Center of SB RAS, Irkutsk: ISDCT SB
RAS; \url{http://hpc.icc.ru}, accessed 16.05.2019).
\end{acks}

\begin{acks}[Disclosure of Potential Conflicts of Interest]
 The authors declare that they have no conflicts of interest.
\end{acks}

 \bibliographystyle{spr-mp-sola}
 
\end{article}
\end{document}